\documentclass[a4paper,12pt]{article}
\usepackage{cite}
\usepackage{amssymb}
\usepackage{amsmath}
\usepackage{amsfonts}
\usepackage{amsthm}
\usepackage{mathtools}
\usepackage[utf8]{inputenc}
\usepackage{color}
\usepackage{graphicx}
\usepackage{color}
\usepackage[active]{srcltx}

\usepackage{here}
\usepackage{cite}
\usepackage{soul}
\usepackage[affil-it]{authblk}

\usepackage{titlesec}
\titleformat{\subsection}
       {\normalfont\fontsize{13}{17}\itshape}{\thesubsection}{1em}{}

\input paperdef

\begin{document}
\null\hfill CERN-TH-2023-124

\thispagestyle{empty}

\title{
Split NMSSM from dimensional reduction of a $10D$, $\mathcal{N}=1$ $E_8$ over $SU(3)/U(1)\times U(1)\times Z_3$}
\date{}
\author{Gregory Patellis$^{1}$\thanks{email: grigorios.patellis@tecnico.ulisboa.pt}~, 
Werner Porod$^2$\thanks{email: porod@physik.uni-wuerzburg.de}~
and George Zoupanos$^{3,4,5}$\thanks{email: george.zoupanos@cern.ch}\\
{\small
$^1$ Centro de Física Teórica de Partículas - CFTP, Departamento de Física,\\ Instituto Superior Técnico, Universidade de Lisboa,\\
Avenida Rovisco Pais 1, 1049-001 Lisboa, Portugal \\
$^2$ Institut für Theoretische Physik und Astrophysik, University of Würzburg, \\ Emil-Hilb-Weg 22, D-97074 Würzburg, Germany \\
$^3$ Physics Department,   National Technical University, 157 80 Zografou, Athens, Greece\\
$^4$ Max-Planck Institut f\"ur Physik, F\"ohringer Ring 6, D-80805
  M\"unchen, Germany \\
$^5$ Theoretical Physics Department, CERN, Geneva, Switzerland \\
}
}

{\let\newpage\relax\maketitle}

\begin{abstract}

We examine an extension of the Standard Model which results from a $10D$, $\mathcal{N}=1$, $E_8$ gauge theory. The theory is dimensionally reduced over a $M_4 \times B_0/ \mathbf{Z}_3 $ space, where $B_0$ is the nearly-K\"ahler manifold
$SU(3)/U(1) \times U(1)$ and $\mathbf{Z}_3$ is a freely acting discrete group on $B_0$ that triggers a Wilson flux breaking, leading to an $\mathcal{N}=1$,  $SU(3)^3\times U(1)^2$ effective theory in $4D$.
At lower energies we are left with the Split NMSSM. Its 2-loop analysis yields third generation quark and light Higgs masses within the experimental limits and predicts a neutralino LSP mass $<800$~GeV.
\end{abstract}


\section{Introduction}\label{intro}
For over half a century, the unification of all fundamental interactions has been a primary objective for theoretical physicists. The scientific community has placed great importance on this pursuit, leading to the development of several intriguing approaches over the past decades. Among these approaches, those that incorporate the existence of extra dimensions have garnered significant attention and prominence.
The concept of extra dimensions finds strong support within the framework of superstring theories. Among these theories, the heterotic string \cite{Gross:1985fr}, defined in ten dimensions, stands out as particularly promising due to its potential for experimental testability. The phenomenological aspects of the heterotic string are especially notable in the resulting Grand Unified Theories (GUTs), which contain the gauge group of the Standard Model. These GUTs are obtained through a process involving the compactification of the $10D$ spacetime and dimensional reduction of the initial $E_8 \times E_8$ gauge theory.
Interestingly, even before the formulation of superstring theories, an alternative framework emerged that focused on the dimensional reduction of higher-dimensional gauge theories. This provided another avenue for exploring the unification of fundamental interactions. The endeavor to unify fundamental interactions, which shared common objectives with many of the superstring theories, was first investigated by Forgacs-Manton (F-M) and Scherk-Schwartz (S-S). F-M explored the concept of Coset Space Dimensional Reduction (CSDR) \cite{Forgacs:1979zs,Kapetanakis:1992hf, Kubyshin:1989vd}, while S-S focused on the group manifold reduction \cite{Scherk:1979zr}. These pioneering works delved into the reduction of dimensions and the implications for unification, laying the foundation for further developments in the field, including the present study.

In both the S-S and F-M mechanisms, the gauge and scalar sectors are unified in the higher-dimensional theory. Specifically in the case of CSDR, fermions from the higher-dimensional theory result in Yukawa interactions in the $4D$ one. Furthermore, for certain numbers of total dimensions, if the initial theory is $\mathcal{N}=1$ supersymmetric, one can start with just a vector supermultiplet which accommodates both gauge and fermionic fields, thus unifying the field content even more. Additionally, it is noteworthy that this process can lead to the emergence of $4D$ chiral theories \cite{Manton:1981es}. 
The convergence of the two approaches can be observed through CSDR, which incorporates the predictions of the heterotic string such as the number of additional dimensions and the gauge group of the initial theory  (although it should be noted that the S-S does not lead to chiral fermions in $4D$ since the dimensional reduction is performed over a group manifold).
Given that superstring theories are consistent in $10D$, it is necessary to distinguish the extra dimensions from the four observable dimensions using a metric compactification before determining the resulting $4D$ theory. Then a suitable compactification manifold choice could result in $4D$ $\mathcal{N}=1$ supersymmetric theories that aim for realistic GUTs.

In the process of reducing the dimensions of an $\mathcal{N}=1$ supersymmetric gauge theory, a crucial and desirable property is the preservation of the original theory's amount of supersymmetry in the resulting $4D$ theory. By imposing this requirement, the compact internal Calabi-Yau (CY) manifolds emerge as highly promising candidates \cite{Candelas:1985en}. Nevertheless, due to the emergence of the moduli stabilization problem, there has been a shift towards exploring flux compactification in a broader range of internal spaces, specifically those with $SU(3)$-structure. In these types of manifolds, one considers a background non-vanishing, globally defined spinor. This spinor remains covariantly constant wrt a connection including torsion, whereas in the case of Calabi-Yau manifolds, this property holds for a Levi–Civita connection. 
Under this light, the class of nearly-Kähler manifolds of the SU(3)-structure manifolds was considered \cite{LopesCardoso:2002vpf,Strominger:1986uh,Lust:1986ix,Castellani:1986rg,Becker:2003yv,Becker:2003sh,butruille2006homogeneous}.
In particular the class of $6D$ homogeneous nearly-Kähler manifolds (that also admit a connection with torsion) consists of the non-symmetric coset spaces $G_2/SU(3)$, $Sp(4)/SU(2)\times U(1)_{non-max}$, $SU(3)/U(1)\times U(1)$ and the group manifold $SU(2)\times SU(2)$ \cite{Manousselis:2005xa,Chatzistavrakidis:2008ii,Chatzistavrakidis:2009mh,Klaput:2011mz}.  A notable characteristic of CSDR, in contrast to the case of CY, is its ability to lead to $4D$ theories that include soft supersymmetry breaking terms when reducing a $10D$, $\mathcal{N}=1$ theory over a non-symmetric coset space \cite{Manousselis:2001xb,Manousselis:2000aj,Manousselis:2004xd} (while it breaks supersymmetry completely when the theory is reduced over symmetric cosets).

In our research, we focus on the dimensional reduction of an N=1 supersymmetric gauge theory and emphasize the preservation of supersymmetry in the resulting $4D$ theory. This article revisits the dimensional reduction of $E_8$ over the modified flag manifold $SU(3)/U(1)\times U(1)\times\mathbf{Z}_3$, where the latter represents the non-symmetric coset space $SU(3)/U(1)\times U(1)$ with the addition of the freely acting discrete symmetry $\mathbf{Z}_3$. This specific choice allows for the induction of the Wilson flux breaking mechanism, enabling further reduction of the gauge symmetry of the $4D$ GUT to $SU(3)^3$, along with two U(1) symmetries \cite{Kapetanakis:1992hf,Manousselis:2001xb,Chatzistavrakidis:2008ii,Irges:2011de} (see also \cite{Lust:1985be}).
Upon reduction, the potential of the resulting $4D$ theory exhibits terms that can be identified as soft breaking terms. This indicates that the resulting theory represents a softly broken N=1 supersymmetric theory.

In the case examined the radii are chosen to be small, such that the compactification matches the unification scale, and, since the soft terms are of geometrical origin, all sfermions become super heavy and eventually decouple (an older version of this model with different choice of radii can be found in \cite{Manolakos:2020cco} and \cite{Patellis:2020cue}). All exotic trinification fields decouple as  well and due to a specific choice of radii we are led to a Split version \cite{Giudice:2004tc} of the  Next-to-Minimal Supersymmetric Standard Model (NMSSM) \cite{Ellwanger:2009dp} (for the Split NMSSM see \cite{Demidov:2006zz,Gabelmann:2019jvz}), in which a singlet field couples to the Higgs field and the lighter supersymmetric particles acquire masses lower than 1 TeV.

This article is organised as follows. In \refse{csdr} we review the basics of the CSDR and describe the reduction of the $10D$, $\mathcal{N}=1$, $E_8$ gauge theory over the modified flag manifold. In \refse{wilson} we apply the breaking by Wilson fluxes and describe the $4D$ effective trinification model. \refse{GUTbreaking} presents key information about the $4D$ theory's parameters, terms and breaking and explains how the effective split NMSSM theory emerges below breaking. The effective theory and its 2-loop analysis and results are presented in \refse{SplitNMSSM} and we conclude in \refse{conclusions} with some final remarks.

\section{Dimensional Reduction of $E_8$ over $SU(3)/U(1)\times U(1)$}\label{csdr}

Let us start with a brief review of the CSDR mechanism. For a comprehensive understanding of the geometric aspects related to coset spaces, we recommend \cite{Kapetanakis:1992hf,Castellani:1999fz}, while a thorough examination of the fundamental elements of CSDR, including the generalized methodology of reduction and the treatment of constraints can be found in \cite{Kapetanakis:1992hf}. 

Consider a $d$-dimensional coset $S/R$, where $S$ is a Lie group and $R$ its subgroup ($d=dimS-dimR$), on which the extra dimensions of the space $M^4\times S/R$ are compactified. The $S$ group acts as a symmetry group on the extra coordinates. The central idea of the CSDR scheme is that an $S$-transformation of the extra coordinates is \textit{compensated} by a gauge transformation of the fields that are defined on the $D$-dimensional space, where $D=d+4$. Therefore, a gauge invariant Lagrangian on this space is independent of the extra coordinates. Fields that are defined in such a way are called symmetric. 

Consider now a $D$-dimensional Yang-Mills-Dirac theory with a gauge group $G$, which is defined on a manifold $M^D$ compactified on $M^4\times S/R$. Its action will then be given by
\begin{align}
S&=\int d^4xd^dy\sqrt{-g}\left[-\frac{1}{4}\text{Tr}(F_{MN}F_{K\Lambda})g^{MK}g^{N\Lambda}+\frac{i}{2}\bar{\psi} \Gamma^MD_{M}\psi\right]\,,\label{actioncsdr}
\end{align}
where $D_M$ is the covariant derivative expressed as
\begin{equation}
    D_M=\partial_M-\theta_M-igA_M\,, \,\text{with}\,\,\, \theta_M=\frac{1}{2}\theta_{MN\Lambda}\Sigma^{N\Lambda}\,
\end{equation}
the spin connection and  $F_{MN}$  the field strength~tensor of $A_M$:
\begin{equation}
    F_{MN}=\partial_M A_N-\partial_NA_M-ig[A_M,A_N]\,.
\end{equation}
The $A_{\mu}$ and $\psi$ fields are symmetric, where $\psi$ is a spinor representing the theory's fermions and is accommodated in the $F$ representation of $G$.

We denote the Killing~vectors of $S/R$ as $\xi^{\alpha}_{A}$ ($A=1,\ldots,\text{dim}S$ and $\alpha=\text{dim}R+1,\ldots,\text{dim}S$). Consider also a gauge transformation, $W_A$ related to the Killing vector~$\xi_A$. The condition that all kinds of fields (scalars, vectors and spinors) that live on the coset~space are symmetric 
translates to the constraints:
\begin{align}
\delta_A\phi&=\xi^{~\alpha}_{A}\partial_\alpha\phi=D(W_A)\phi, \nonumber\\
\delta_AA_{\alpha}&=\xi^{~\beta}_{A}\partial_{\beta}A_{\alpha}+\partial_{\alpha}\xi^{~\beta}_{A}A_{\beta}=\partial_{\alpha}W_A-[W_A,A_{\alpha}], \label{constraints}\\
\delta_A\psi&=\xi^{~\alpha}_{A}\partial_\alpha\psi-\frac{1}{2}G_{Abc}\Sigma^{bc}\psi=D(W_A)\psi\,,\nonumber
\end{align}
where $D(W_A)$ is the transformation $W_A$ in the appropriate representation, according to the representation in which each field is assigned and only depends on the coordinates of the coset. The solutions of the constraints of \eqref{constraints} determine both the gauge group and the surviving field content of the reduced $4D$ theory. In particular, the solution of the first constraint makes clear that the gauge group of the $4D$ theory is $H=C_G(R_G)$, which is the centralizer of $R$ in $G$, and also to the fact that the first part of $A_M=(A_{\mu},A_{\alpha})$ is identified as the $4D$ gauge field and has no dependence on the coset coordinates. Then $A_{\alpha}$ is identified as a set of scalar fields in the $4D$ theory that assume the role of intertwining operators among the representations of $R$~in $G$ and in $S$. This is a results of the second constraint. The decomposition of $G$ and the decomposition of $S$ under $R$ give an insight on the representation of the produced scalars in the $4D$ theory:
\begin{equation}
G\supset R_G\times H\,,\qquad
\mathrm{adj}G=(\mathrm{adj}R,1)+(1,\mathrm{adj}H)+\sum(r_i,h_i)\,,
\end{equation}
\begin{equation}
S\supset R\,,\qquad
\mathrm{adj}S=\mathrm{adj}R+\sum s_i\,.
\end{equation}
The surviving scalar spectrum in the four-dimensional theory is determined by examining the irreducible representations $r_i$ and $s_i$ of $R$. When the representations $r_i$ and $s_i$ are identical, the representation $h_i$ of $H$ corresponds to a scalar multiplet. Any additional scalars that are not accounted for in $h_i$ are projected out and do not appear in the $4D$ theory. This selection process allows us to determine the surviving $4D$ scalar spectrum.

The third constraint in  \eqref{constraints} is associated with the spinorial component and its form and characteristics in the $4D$ theory. The approach is similar to what was described above regarding the scalar fields. By solving the constraint, it becomes evident that the $4D$ spinors solely depend on the $4D$ coordinates. Moreover, like the scalar fields, they relate the induced representations of $R$ between $SO(d)$ and $G$ as intertwining operators. The representation $f_i$ of the $4D$ gauge group $H$ to which fermions are assigned is determined by the decomposition rule of the representation $F$ of $G$ wrt $R_G\times H$ and the spinorial representation of $SO(d)$ under $R$:\small
\begin{equation}
 G\supset R_G\times H\,,\qquad F=\sum (r_i,f_i),
\end{equation}
\begin{equation}
SO(d)\supset R\,,\qquad \sigma_d=\sum \sigma _j\,.
\end{equation}\normalsize
In a similar manner to the scalar case discussed above, the representations $r_i$ and $\sigma_i$ are effectively identical. For each pair of these representations, an $f_i$ multiplet of spinors survives in the $4D$ theory. 

It is important at this point to comment on the nature of fermions before and after dimensional reduction. If the fermions introduced in the higher-dimensional theory are assumed to be Dirac fermions, the surviving $4D$ fermions will not be chiral. This is certainly an unwelcome property. However, starting  on an even $D$-dimensional spacetime and imposing the Weyl condition, the resulting $4D$ theory will have chiral fermions. Moreover, fermions assigned to the adjoint representation of the gauge group in a higher-dimensional theory defined on a spacetime with $D=2n+2$ dimensions, satisfying the Weyl condition, lead to two sets of chiral $4D$ fermions with identical quantum numbers of the components of the two sets. If the condition imposed on the fermions of the higher-dimensional theory also includes the Majorana condition, the doubling of the fermionic spectrum does not occur in $4D$. Both the Weyl and Majorana conditions can be imposed if $D=4n+2$.\\

We now turn or focus towards the practical implementation of the CSDR scheme and the case at hand. We start from an $\mathcal{N}=1$ supersymmetric $E_8$ theory in $10D$ with a single vector representation and Weyl-Majorana fermions. Following all the above-mentioned ruleset, it is dimensionally reduced  over the non symmetric space $SU(3)/U(1)\times U(1)$ \cite{Kapetanakis:1992hf,Manousselis:2001xb,Lust:1985be}. The $4D$ action that remains after the reduction is:
\begin{align}
    S=C\int d^4x\,\mathrm{tr}\left[-\frac{1}{8}F_{\mu\nu}F^{\mu\nu}-\frac{1}{4}(D_\mu\phi_a)(D^\mu\phi^a)\right]+V(\phi)+\frac{i}{2}\bar{\psi}\Gamma^\mu D_\mu\psi-\frac{i}{2}\bar{\psi}\Gamma^aD_a\psi \,,\label{4Daction}
\end{align}
where the scalar potential (still in formal form since the constraints of \eqref{constraints} have not yet been applied) is given by
\begin{align}
    V(\phi)&=-\frac{1}{4}g^{ac}g^{bd}\mathrm{tr}\left(f_{ab}^{~~C}\phi_C-i[\phi_a,\phi_b])(f_{cd}^{~~D}\phi_D-i[\phi_c,\phi_d]\right)\label{four-dimpotential}
\end{align}
and $\mathrm{tr}(T^iT^j)=2\delta^{ij}$, where $T^i$ are the generators of the gauge~group, $C$ is the volume of the coset,~$D_\mu=\partial_\mu-igA_\mu$ is~the $4D$ covariant derivative, $D_a$ is~that of the coset and~the coset metric is given (in terms~of its radii) by $g_{\alpha\beta}=\text{diag}(R_1^2,R_1^2,R_2^2,R_2^2,R_3^2,R_3^2)$.

The determination of the $4D$ gauge group relies on the specific embedding of the $R=U(1)\times U(1)$ subgroup within $G=E_8$. This gauge group is obtained by considering the centralizer of $R=U(1)\times U(1)$ within $G=E_8$:
\begin{align}
H=C_{E_8}(U(1)_A\times U(1)_B)=E_6\times U(1)_A\times U(1)_B\,.
\end{align}
Upon resolving the constraints, the remaining scalar and fermion fields in the $4D$ theory are obtained through the decomposition of the 248 (adjoint) representation of $E_8$ under $U(1)_A\times U(1)_B$.
To determine the representations of the surviving fields, it is necessary to examine the decomposition of the vector and spinor representations of $SO(6)$ under $R=U(1)_A\times U(1)_B$ (following the discussion above).

As a result, the surviving gauge fields, belonging to $E_6\times U(1)_A\times U(1)_B$, are accommodated within three $\mathcal{N}=1$ vector supermultiplets, while the matter fields  are organized into six chiral multiplets. Among these, three are $E_6$ singlets carrying $U(1)_A\times U(1)_B$ charges, while the remaining chiral multiplets transform under the $E_6\times U(1)_A\times U(1)_B$ symmetry. The unconstrained matter fields are:
\begin{align*}
A_i \sim 27_{(3,\frac{1}{2})}, \quad  B_i \sim
27_{(-3,\frac{1}{2})}, \quad  \Gamma_i \sim 27_{(0,-1)}, \quad
A \sim 1_{(3,\frac{1}{2})}, \quad  B \sim
1_{(-3,\frac{1}{2})}, \quad  \Gamma \sim 1_{(0,-1)}
\end{align*}
and the scalar potential becomes:
\begin{align*} 
V(\alpha^i,\alpha,\beta^i,\beta,\gamma^i,\gamma)=& \frac{g^2}{2}\bigg[\frac{2}{5}\left(\frac{1}{R_1^4}+\frac{1}{R_2^4}+\frac{1}{R_3^4}\right)\nonumber\\
&+\bigg(\frac{4R_1^2}{R_2^2 R_3^2}-\frac{8}{R_1^2}\bigg)\alpha^i \alpha_i + \bigg(\frac{4R_1^2}{R_2^2 R_3^2}- \frac{8}{R_1^2}\bigg)\bar{\alpha} \alpha \nonumber\\
&+\bigg(\frac{4R_2^2}{R_1^2 R_3^2}-\frac{8}{R_2^2}\bigg)\beta^i \beta_i +\bigg(\frac{4R_2^2}{R_1^2 R_3^2}-\frac{8}{R_2^2}\bigg)\bar{\beta} \beta \nonumber\\
&+\bigg(\frac{4R_3^2}{R_1^2 R_2^2}-\frac{8}{R_3^2}\bigg)\gamma^i \gamma_i +\bigg(\frac{4R_3^2}{R_1^2 R_2^2}-\frac{8}{R_3^2}\bigg)\bar{\gamma} \gamma \nonumber\\
&+\bigg[80\sqrt{2} \bigg(\frac{R_1}{R_2 R_3}+\frac{R_2}{R_1 R_3}+\frac{R_3}{R_2 R_1}\bigg)d_{ijk}\alpha^i \beta^j \gamma^k\nonumber\\
&+80\sqrt{2}\bigg(\frac{R_1}{R_2 R_3}+\frac{R_2}{R_1 R_3}+\frac{R_3}{R_2 R_1}\bigg)\alpha \beta \gamma +h.c \bigg] \nonumber \\
&+ \frac{1}{6}\bigg( \alpha^i(G^\alpha)_i^j\alpha_j+\beta^i(G^\alpha)_i^j\beta_j+\gamma^i(G^\alpha)_i^j\gamma_j\bigg)^2 \nonumber\\
&+\frac{10}{6}\bigg( \alpha^i(3\delta_i^j)\alpha_j+\bar{\alpha}(3)\alpha+\beta^i(-3\delta_i^j)\beta_j+\bar{\beta}(-3)\beta \bigg)^2 \nonumber\\
&+\frac{40}{6}\bigg( \alpha^i(\tfrac{1}{2}\delta_i^j)\alpha_j+\bar{\alpha}(\tfrac{1}{2})\alpha+\beta^i(\tfrac{1}{2}\delta_i^j)\beta_j+\bar{\beta}(\tfrac{1}{2})\beta+\gamma^i(-1\delta_i^j)\gamma^j+\bar{\gamma}(-1)\gamma \bigg)^2 \nonumber
\end{align*}
\begin{align} 
&+40\alpha^i \beta^j d_{ijk}d^{klm} \alpha_l \beta_m+40\beta^i
\gamma^j d_{ijk}d^{klm} \beta_l
\gamma_m+40 \alpha^i \gamma^jd_{ijk} d^{klm} \alpha_l \gamma_m \nonumber\\
&+40(\bar{\alpha}\bar{\beta})(\alpha\beta)+40(\bar{\beta}\bar{\gamma})(\beta\gamma)+40(\bar{\gamma}\bar{\alpha})(\gamma
\alpha)\bigg]\,,\label{E6_pot}
\end{align}
and is positive~definite, while $\alpha^i,\alpha,\beta^i,\beta,\gamma^i,\gamma$ are the scalar components of $A^i,B^i,\Gamma^i$ and $A,B,\Gamma$. In the above expression of the scalar potential,~the $F-, D-$ and soft supersymmetry~breaking terms are~identified. The $F$-terms (last two lines of \refeq{E6_pot}) emerge from~the superpotential:
\begin{align}
\mathcal{W}(A^i,B^j,\Gamma^k,A,B,\Gamma)=\sqrt{40}d_{ijk}A^iB^j\Gamma^k+\sqrt{40}AB\Gamma\,,
\end{align}
The $D$-terms are identified (lines 7-9 of \refeq{E6_pot}) in their usual structure:
\begin{align}
\frac{1}{2}D^{\alpha}D^{\alpha}+\frac{1}{2}D_1D_1+\frac{1}{2}D_2D_2
\end{align}
and the remaining terms of \refeq{E6_pot} (save the constant first term) admit the interpretation~of soft scalar masses and trilinear soft terms. 
The gaugino mass, $M$,  is also of geometrical origin, although it demonstrates different behaviour compared to the soft masses:
\begin{equation}
    M=(1+3\tau)\frac{R_1^2+R_2^2+R_3^2}{8\sqrt{R_1^2R_2^2R_3^2}}\,.\label{gaugino}
\end{equation}
This shows that the gauginos would naturally obtain mass at the compactification scale \cite{Kapetanakis:1992hf}. This, however, is prevented by the inclusion of the (con)torsion \cite{Manousselis:2001xb}, which will be appropriately in order for the following model to feature a electroweak (EW) scale unified gaugino mass.

\section{Wilson Flux and the Effective Unified Theory} \label{wilson}

In the preceding section, we discussed the application of the CSDR on a higher-dimensional $E_8$ gauge theory, which dimensionally reduced over an $SU(3)/U(1)\times U(1)$ coset space, resulting in a $4D$ $E_6\times U(1)_A \times U(1)_B$ gauge theory. However, it should be noted that the $27$ multiplet is insufficient to break the $E_6$ gauge symmetry. To achieve the desired reduction of the gauge symmetry, the Wilson flux breaking mechanism is employed \cite{Kozimirov:1989kn, Zoupanos:1987wj, Hosotani:1983xw}.

We will briefly review review the basics of the Wilson flux
mechanism here. In the previous section the dimensional reduction was performed over a simply connected manifold, specifically a coset space $B_0=S/R$. However, there is an alternative option where the manifold can be chosen to be multiply connected. This is achieved by considering $B=B_0/F^{S/R}$, where $F^{S/R}$ is a freely-acting discrete symmetry of the manifold $B_0$. For each element $g\in F^{S/R}$, there is a corresponding element $U_g$ in the $4D$ gauge group $H$, which may be viewed as the Wilson loop:
\begin{equation}
U_g={\mathcal{P}}exp\left(-i\oint_{\gamma_g} T^a A^{~a}_M dx^M \right),
\end{equation}
where $A^{~a}_M$ are the gauge fields, $\gamma_g$ a contour representing the element $g$ of
$F^{S/R}$, $T^a$  are the generators of the group and $\mathcal{P}$  denotes the path ordering. In the case where the considered manifold is simply connected, the vanishing of the field strength tensor implies that the gauge field can be set to zero through a gauge transformation. However, when $\gamma_g$ is chosen to be non-contractible to a point, we have $U[\gamma]\neq 1$, and the gauge field cannot be gauged away. This means that the vacuum field strength does not lead to $U_g=1$. 
As a result, a homomorphism of $F^{S/R}$ into~$H$ is induced with an image $T^H$, which is the subgroup of $H$ generated by $U_g$. Furthermore, consider a field $f(x)$ defined on $B_0$. It is evident that $f(x)$ is equivalent to another field on $B_0$ that satisfies $f(g(x))=f(x)$ for every $g\in F^{S/R}$. The presence of $H$ generalizes this statement:
\begin{align}
f(g(x))=U_gf(x)\,.\label{generalisedstatement}
\end{align}
Regarding the gauge symmetry that remains by the vacuum, in the vacuum state it is given that $A_\mu^a=0$, and consider also a gauge transformation by the coordinate-dependent matrix $V(x)$ of $H$. In order to keep $A_\mu^{~a}=0$ and preserve the vacuum invariance, the matrix $V(x)$ must be chosen to be constant. Additionally, $f\to Vf$ is consistent with \refeq{generalisedstatement} only if
\begin{equation}
 [V,U_g]=0\, \label{consistenconditionforwilson}    
\end{equation}
for every~$g\in F^{S/R}$. Hence, the subgroup of $H$ that remains unbroken is the centralizer of $T^H$ in $H$. As for the matter fields that survive in the theory, meaning the matter fields that satisfy the condition in \refeq{generalisedstatement}, they must be invariant~under the combination:
\begin{align*}
F^{S/R}\oplus T^H.
\end{align*}
The freely-acting discrete symmetries, $F^{S/R}$, of $B_0=S/R$ are the center of $S$, $Z(S)$ and $W=W_S/W_R$, where it is understood that $W_S$~and $W_R$ are the Weyl groups of $S$ and $R$, respectively. In the present article we have
\begin{equation}
F^{S/R}=\mathbb{Z}_3  \subseteq W = S_{3},
\end{equation}
since the coset space is selected to be $B_0=SU(3)/U(1)\times U(1)$.\\

In other words, with the implementation of the Wilson flux breaking mechanism the theory is projected in a manner that the fields that survive are the ones that remain invariant under the action of the freely acting discrete symmetry $\mathbf{Z}_3$  on their gauge and geometric indices. In the case studied, the $\mathbf{Z}_3$'s non-trivial action on the gauge indices of the fields is parameterized by the matrix
\cite{Chatzistavrakidis:2010xi}:
\begin{equation}
    \gamma_3=\text{diag}\{\mathbf{1}_3,\omega \mathbf{1}_3, \omega^2
\mathbf{1}_3\}\,,
\end{equation}
where $\omega=e^{i\frac{2\pi}{3}}$ is the phase that acts on the gauge fields of the gauge~theory.  The surviving gauge fields are the ones that satisfy the condition:
\begin{equation}
[A_M,\gamma_3]=0\;\;\Rightarrow A_M=\gamma_3 A_M \gamma_3^{-1}\label{filteringgaugefields}
\end{equation}
and the remaining~gauge symmetry is $SU(3)_c\times SU(3)_L \times SU(3)_R\times U(1)_A\times U(1)_B$. The  abelian $U(1)$s are the $R$-symmetry of the theory, which is closely  interrelated to supersymmetry.
The matter~counterpart of \refeq{filteringgaugefields} is:
\begin{equation}
A^i=\gamma_3A^i,\;\;B^i=\omega
\gamma_3B^i,\;\;\Gamma^i=\omega^2 \gamma_3\Gamma^i\,,\qquad A= A,\;\;B=\omega B,\;\;\Gamma=\omega^2 \Gamma\,.
\end{equation}\label{wilsonprojectionmatter}
The representations of the ${SU(3)_c\times SU(3)_L\times SU(3)_R}$ part of the gauge group, in which~the above fields belong, can be obtained after examining the decomposition of the 27 representation of $E_6$ under the trinification group, $(1,3,\bar 3)\oplus(\bar3,1,3)\oplus(3,\bar 3,1)$.
Therefore, the projected theory has the following matter content:
\begin{align*}
A_1&\equiv L\sim(\mathbf{1},\mathbf{3},\overline{\mathbf{3}})_{(3,1/2)},~~~~~~B_3\equiv q^c\sim(\overline{\mathbf{3}},\mathbf{1},\mathbf{3})_{(-3,1/2)},~~~~~~\Gamma_2\equiv Q\sim(\mathbf{3},\overline{\mathbf{3}},\mathbf{1})_{(0,-1)}\\
A&\equiv\theta\sim(\mathbf{1},\mathbf{1},\mathbf{1})_{(3,1/2)}
\end{align*}
where the former three constitute the remaining components of $A^i,B^i,\Gamma^i$. Collectively, they form a 27 representation of $E_6$, which corresponds to the content representing one generation in the surviving model.
To achieve a spectrum comprising three generations, it is possible to introduce non-trivial monopole charges in the $U(1)$s in $R$. This leads to the existence of three replicas of the aforementioned fields, resulting in a total of three sets of fields \cite{Dolan:2003bj}. This  allows for the generation of a spectrum consisting of three generations. The trinification multiplets $L,q^c,Q$ can be assigned and written in a more illuminating way:
\begin{eqnarray*}L^{(l)}=\left(\begin{array}{ccc}
 H_d^0 & H_u^+ & \nu_L \\
 H_d^- & H_u^0 & e_L \\
 \nu^c_R & e^c_R & N
 \end{array}\right)\,,\,\,
  q^{c(l)}=\left(\begin{array}{ccc}
 d^{c1}_R & u^{c1}_R & D^{c1}_R \\
 d^{c2}_R & u^{c2}_R & D^{c2}_R \\
 d^{c3}_R & u^{c3}_R & D^{c3}_R
 \end{array}
 \right)\,,\,\, Q^{(l)}=\left(\begin{array}{ccc}
 d^1_L & d^2_L & d^3_L \\
 u^1_L & u^2_L & u^3_L \\
 D^1_L & D^2_L & D^3_L
 \end{array}\right)\,,
\end{eqnarray*}
where the index $l=1,2,3$ denotes the generation. It now becomes clear that $q^c$ and $Q$ are quark multiplets, while $L$ accommodates both the lepton and the Higgs sector. The quark multiplets also contain the vector-like down-type quarks $D^{(l)}$, which will eventually be  $SU(2)_L$ singlets, while $L$ also features the right-handed quarks $\nu_R^{c(l)}$ and the sterile neutrino-like fields $N^{(l)}$. It is useful to note that there are three generations of Higgs doublets. Finally, there are three trinification singlets, $\theta^{(l)}$.\\

At this point it is useful to consider that if an effective $4D$ theory is renormalizable by power counting, then it is consistent to consider it a renormalizable theory \cite{Polchinski:1983gv}. Under this light, we respect all the symmetries and the model structure that are derived by the higher-dimensional theory and its dimensional reduction. We treat, however, all the parameters of the effective theory as  free  parameters, to the extent allowed by symmetries. In particular, all kinetic terms and $D$-terms of \refeq{4Daction} have the gauge coupling, $g$, as dictated by the gauge symmetry of the model. All superpotential terms must also share the same coupling, in order to respect supersymmetry. The freedom given by this treatment becomes apparent in the soft sector, where each term is allowed to its own coupling.

Taking all the above into account, the superpotential of the $SU(3)_c\times SU(3)_L\times SU(3)_R\times U(1)_A\times U(1)_B$ effective theory will be:
\begin{align}
\mathcal{W}^{(l)}=C^{(l)}d^{abc}L_a^{(l)}q_b^{c(l)}Q_c^{(l)}\,,\label{superpot}
\end{align}
since the $B$ and $\Gamma$ trinfication singlets were projected out by the Wilson fluxes. using the same arguments, the soft sector of the scalar potential is now given by:
\begin{align}
V_{soft}^{(l)}=&\left(\frac{c_{L_1}^{(l)}R_1^2}{R_2^2R_3^2}-\frac{c_{L_2}^{(l)}}{R_1^2}\right)\big<L^{(l)}|L^{(l)}\big>+\left(\frac{c_{\theta_1}^{(l)}R_1^2}{R_2^2R_3^2}-\frac{c_{\theta_1}^{(l)}}{R_1^2}\right)|\theta^{(l)}|^2\nonumber \\
&+\left(\frac{c_{q_1^c}^{(l)}R_2^2}{R_1^2R_3^2}-\frac{c_{q_2^c}^{(l)}}{R_2^2}\right)\big<q^{c(l)}|q^{c(l)}\big>
+\left(\frac{c_{Q_1}^{(l)}R_3^2}{R_1^2R_2^2}-\frac{c_{Q_1}^{(l)}}{R_3^2}\right)\big<Q^{(l)}|Q^{(l)}\big>\nonumber \\
&+\left(\frac{R_1}{R_2R_3}+\frac{R_2}{R_1R_3}+\frac{R_3}{R_1R_2}\right)(c_{\alpha}^{(l)}d^{abc}L_a^{(l)}q_b^{c(l)}Q_c^{(l)}+c_{b}^{(l)}d^{abc}L_a^{(l)}L_b^{(l)}L_c^{(l)}+h.c)\nonumber\\
=& m_{L^{(l)}}^2\big<L^{(l)}|L^{(l)}\big>+m_{q^{c(l)}}^2\big<q^{c(l)}|q^{c(l)}\big>+m_{Q^{(l)}}^2\big<Q^{(l)}|Q^{(l)}\big>+m_{\theta^{(l)}}^2|\theta^{(l)}|^2\nonumber \\ &+(\alpha^{(l)abc}L_a^{(l)}q_b^{c(l)}Q_c^{(l)}+b^{(l)abc}L_a^{(l)}L_b^{(l)}L_c^{(l)}+h.c)\,,\label{soft}
\end{align}
where $c_i^{(l)}$ are free parameters of $\mathcal{O}(1)$ and it is understood that the above equation only involves the scalar components of the denoted superfields. It becomes clear that all sfermions, Higgs bosons and (trificication) singlet scalars acquire a soft mass parameter. Regarding the soft trinilear terms,  soft R-symmetry breaking terms $b_i^{(l)}$ have been taken into account.

\section{Selection of Parameters and GUT Breaking}\label{GUTbreaking}
With the theoretical framework now  fully in place,  we proceed to the specification of the free parameters of the theory, in order to set the scene for the low energy effective theory.

\subsection*{Choice of radii}

We will focus on the case where the compactification scale, $M_C$ is high. In this case Kaluza-Klein modes are irrelevant. Otherwise the eigenvalues of the Dirac and Laplace operators in the $6D$ compactification space would be necessary for our study. In particular, we choose $M_C=M_{GUT}$ and as such we have small radii
\begin{eqnarray*}
R_l\sim \frac{1}{M_{GUT}}~, ~ l=1,2,3 \,.    
\end{eqnarray*}
Since all soft trinilear terms and soft masses are of geometrical origin, one look at \refeq{soft} makes it clear that are of order $M_{GUT}$. This leads us to a split supersymmetric scenario \cite{Giudice:2004tc}. While  $R_2=R_3$ guarantees $m_{q^{c(l)}}^2=m_{Q^{(l)}}^2$, we choose $R_1$ to be \textit{slightly} different than the other two. This, combined with appropriate values of the $c_{\theta_i}^{(l)}$ parameters, provides a 'cancellation' of the $m_{\theta^{(3)}}^2$ mass parameter, which can now be $\sim\mathcal{O}(EW)$.

\subsection*{Further gauge symmetry breaking}\label{breakings}
The spontaneous breaking of the $SU(3)_L\times SU(3)_R\times U(1)_A\times U(1)_B$ can be
triggered by the following vevs:
\[
\langle L_s^{(1)}\rangle=\left(\begin{array}{ccc}
0 & 0&0\\
0&0&0\\
0&0&V_1
\end{array}\right),\;\;
\langle L_s^{(2)}\rangle=\left(\begin{array}{ccc}
0 & 0&0\\
0&0&0\\
V_2&0&0
\end{array}\right)~,\;\;
\langle L_s^{(3)}\rangle=\left(\begin{array}{ccc}
0 & 0&0\\
0&0&0\\
V_3&0&V_4
\end{array}\right)~,
\]
\[
\langle\theta_s^{(1)}\rangle=V_5~,\;\;
\langle\theta_s^{(2)}\rangle=V_6~, 
\]
where the $s$ index denotes the scalar component of each superfield. All these $\mathcal{O}(GUT)$ vevs are singlets under $SU(3)_c$, so they leave the colour group intact. $V_1,V_2$ are necessary to break $SU(3)_L\times SU(3)_R$ \cite{Babu:1985gi}, while one of $V_3,V_4$ and one of $V_5,V_6$ are necessary to break the two $U(1)$s. Following \cite{Giudice:2004tc}, we proceed with the above-mentioned maximal vev content, something that will prove useful in the formation of the low energy model, getting the breaking:
\begin{equation}\label{breaking}
SU(3)_c\times SU(3)_L\times SU(3)_R\times U(1)_A\times U(1)_B  \xrightarrow{V_i} SU(3)_c\times SU(2)_L\times U(1)_Y~.
\end{equation}

\subsection*{Missing terms}\label{missing}

It is obvious that \eqref{superpot} lacks bilinear terms that would serve as $\mu$-terms in the low energy model, since they would violate the two $U(1)$s. For this purpose we take into account the following dim-5 operators \cite{Antoniadis:2008es}  below the unification scale that also respect the abelian symmetries:
\begin{equation}
H_u^{(l)}H_d^{(l)}\overline{\theta}^{(l)}\frac{K^{(l)}}{M}~,\label{trilinearmu}
\end{equation}
where $K^{(l)}$ is the conjugate of any field that acquires a  superheavy vev, namely $N^{(1,3)},\nu_{R}^{(2,3)}$ and $\theta^{(1,2)}$ (or any combination of them, provided the generation index is respected). These terms are non-negligible as $\langle K\rangle/M\sim \mathcal{O}(10^{-1})$. The (natural) generation  diagonality of \refeq{trilinearmu} leads to a very interesting phenomenology. More specifically, since these terms are effectively $\mu$-like terms, the Higgs doublets of the two first generations acquire a superheavy $\mu$ term, since $\langle \theta_s^{(1,2)}\rangle\sim \mathcal{O}(GUT)$, while the term of the third generation survives in the low energy model.  

Similarly, the lepton sector cannot have invariant Yukawa terms. However there are the non-negligible dim-7 operators \cite{Antoniadis:2008ur}: 
\begin{equation}
L^{(l)}\overline{e}^{(l)}H_d^{(l)}\Big(\frac{K^{(l)}}{M}\Big)^3~,
\end{equation}
which allow for lepton masses in each generation. Taking into account the rest of the allowed and non-negligible higher-dim operators, one gets superheavy mass terms for all trinification singlet fields, except $\theta_f^{(3)}$  that gets a $\sim \mathcal{O}(EW)$ mass due to a cancellation among terms.

\section{The Split NMSSM Effective Theory}\label{SplitNMSSM}

We can finally fully sort the particle content left under the $SU(3)_c\times SU(2)_L \times U(1)_Y$ surviving gauge group. The additional fields that usually come with unification, namely the vector-like quarks $D^{(l)}$ and the 'sterile' fields $N^{(l)}$ and $\nu_R^{(l)}$ become supermassive and eventually decouple, something that holds for the first two generations of the (trinification) singlets $\theta^{(1,2)}$ and Higgs doublets $H_{u,d}^{(1,2)}$ (this discussion holds for both fermion and scalar components). 

The torsion is such that the unified gauginos acquire EW masses (of a few TeV) (see \refeq{gaugino}), while all sfermions get superheavy due to the soft masses. 
The soft Higgs mass parameters of the third generation $m_{L^{(3)}}^2\equiv m_{H_{u,d}}^2$ are superheavy, while the last term of \refeq{soft}, in a similar way with \refeq{trilinearmu} contains a soft B-like term for the third generation of Higgs doublets (also for the others, but it is irrelevant for the following)
\begin{equation}
    b^{(3)}H_u^{(3)}\cdot H_d^{(3)} \equiv b H_u\cdot H_d
\end{equation}
is also $\sim\mathcal{O}(GUT)$. Assuming a cancellation between $m_{H_{u,d}}^2$ and $b$ (like in \cite{Gabelmann:2019jvz}), the Higgs doublets and the singlet field of the third generation $H_{u,d}^{(3)}\equiv H_{u,d}$ and $\theta^{(3)}\equiv S$ are light and survive down to the electroweak scale, having an interaction term from \refeq{trilinearmu} (in superfield notation):
\begin{equation}
\lambda S H_u\cdot H_d~\,,
\end{equation}
where we drop the generation indices for the above fields since they are unique in the low energy regime, but also because our analysis will be focused on the third generation of fermions from now on. 

A close look to the above makes it clear that we have a split NMSSM scenario \cite{Demidov:2006zz,Gabelmann:2019jvz} (the notation used here matches \cite{Gabelmann:2019jvz}). In particular, the GUT scale soft Higgs mass parameter causes the heavy Higgs scalars ($H_0,A_H,H^{\pm}$) to beccome superheavy and decouple. Thus, the light scalar sector has three remaining fields, namely the light Higgs boson $h$, the scalar $S$ (which also acquires a  vev) and its CP odd counterpart $A$. Finally, the model has an extra constraint from the higher-energy theory. The top and bottom Yukawa couplings are the same at the unification boundary, due to the structure of the superpotential. Therefore, it is natural to expect a large $tan\beta$.\\

\begin{figure}[H]
\centering
\includegraphics[width=0.496\textwidth]{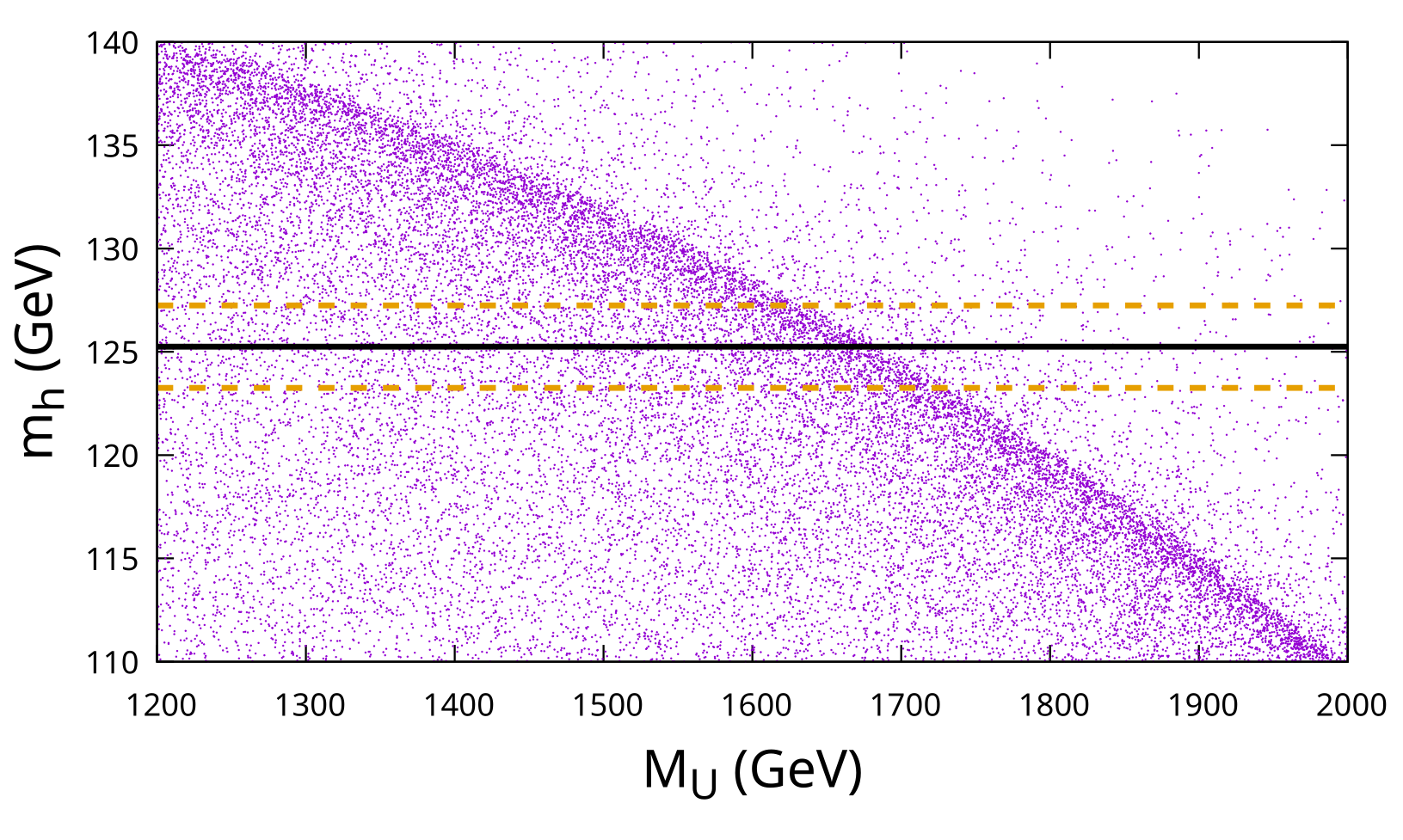}
\includegraphics[width=0.496\textwidth]{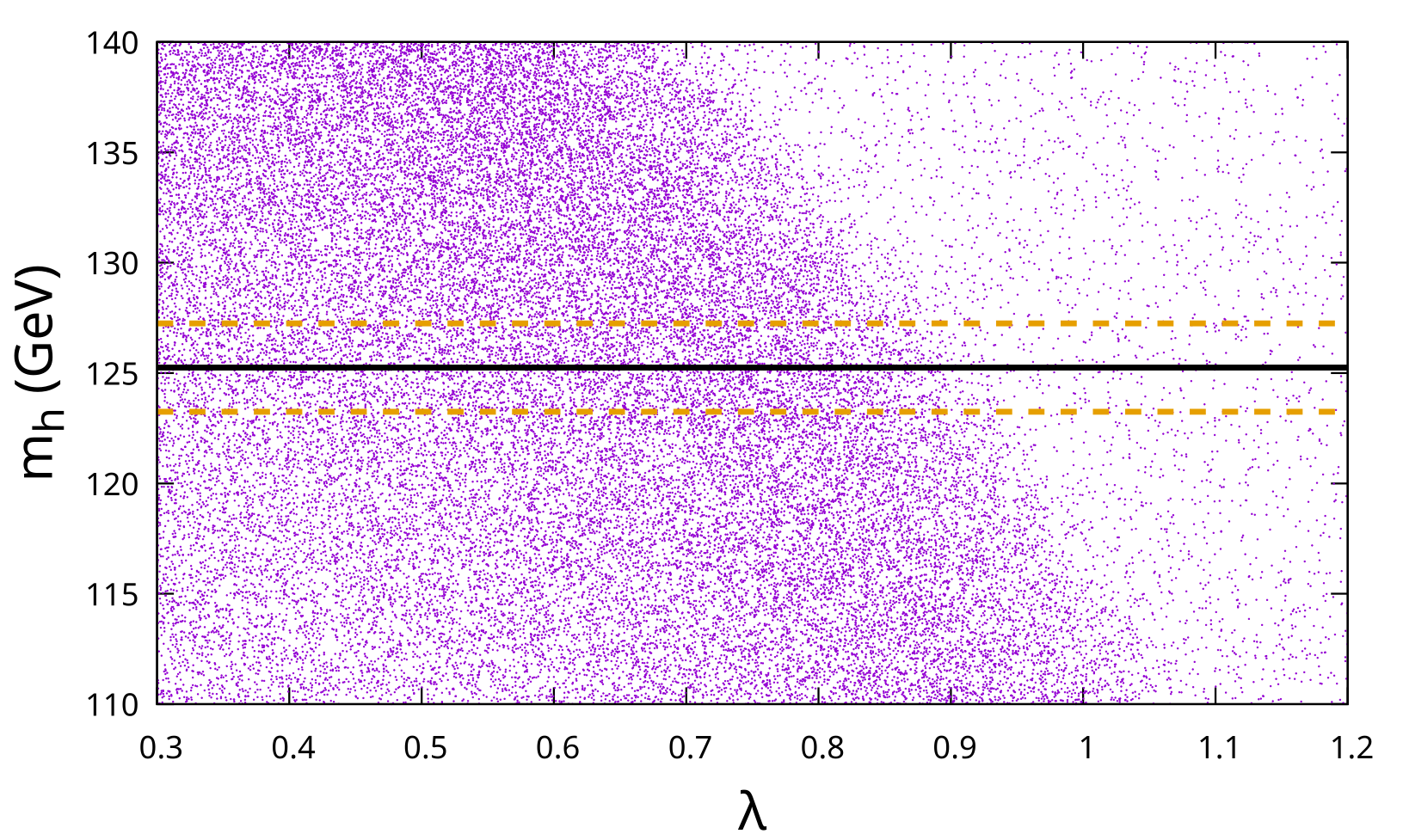}
\caption{\small \textit{Left: the light Higgs boson mass as a function of the unified gaugino mass. Right: the light Higgs boson mass as a function of the trilinear parameter $\lambda$. In both plots the black line denotes the experimental value of the Higgs mass, $m_h=125.25$ GeV, while the orange dashed lines denote the 2~GeV theoretical uncertainties.  
}
}
\label{fig:higgs}
\end{figure}

\begin{figure}[htb!]
\centering
\includegraphics[width=0.8\textwidth]{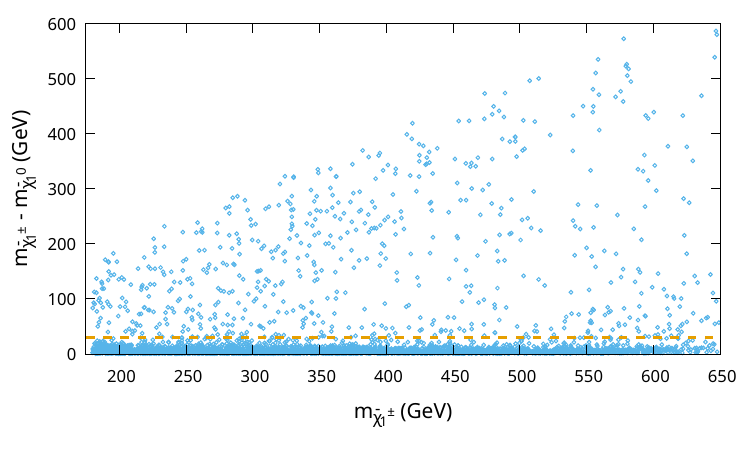}
\caption{\textit{The plot  shows  the mass difference between the lightest chargino and the lightest neutralino, which is the LSP for points that satisfy the Higgs mass theoretical uncertainty of \cite{Slavich:2020zjv}. The orange dashed line denotes the 30~GeV mass difference limit.}
}
\label{fig:charneut}
\end{figure}

In order to produce the (light) particle spectrum of the model, we have implemented the model in \texttt{SARAH}
\cite{Staub:2013tta} and generate a corresponding  \texttt{SPheno} code \cite{Porod:2003um,Porod:2011nf}. At the unification level we use any above-mentioned relations among parameters as boundary conditions for the 2-loop renormalisation group equations (RGEs) that run down to the EW scale.
We also take into account threshold corrections originating from the superheavy particles that decouple and we allow for an extra $5\%$ uncertainty on the boundary condition of the Yukawa couplings. For the following analysis, we use the on-shell in case of the top
quark and the $\overline{\text{MS}}$
in case of the  bottom quark: 
\begin{equation}
m_t=(172.69\pm 0.30)~\text{GeV}~~,~~~~~ m_b(m_b)=(4.18\pm 0.03)~\text{GeV}~~,\label{topbottom}
\end{equation}
as given in \cite{ParticleDataGroup:2022pth}. As expected, in order to satisfy these limits we have $70<\tan\beta<80$. 
Note, that $\beta$ is the angle between $H_u$ and $H^*_d$ which determines the light Higgs doublet at the high scale
at which the second doublet is integrated out.
The light Higgs boson mass, $m_h$, as calculated for our model, is shown in \reffi{fig:higgs} as a function of the unified gaugino mass $M_U$ and the trilinear coupling $\lambda$, where only points that agree with the experimental values of the third generation of quark masses are included. We see that the unified gaugino mass, $M_U$ cannot be more than $1800$GeV, and the most points that satisfy the experimental limits on the Higgs boson mass \cite{ParticleDataGroup:2022pth},
\begin{equation}
m_h^{exp}=(125.25\pm 0.17)~\text{GeV}\,, \label{higgsmass}
\end{equation}
are the ones that have $1600\text{GeV}<M_U<1700\text{GeV}$. 
 We also include a theoretical uncertainty of 2~GeV
\cite{Slavich:2020zjv}.
Similarly, for the trilinear coupling $\lambda$ we see that $\lambda<0.9$ is preferred in order to get the Higgs mass values within the uncertainties.
The lightest neutralino is the lowest supersymmetric particle (LSP). The difference of the lightest chargino mass and the LSP is given in \reffi{fig:charneut} with respect to
mass of the lighest chargino. All points shown have a Higgs mass within the 2~GeV theoretical uncertainty of \cite{Slavich:2020zjv} and also satisfy the lightest chargino mass lower exclusion bounds \cite{ParticleDataGroup:2022pth}. 
Charginos and neutralinos have been
searched for by the LHC experiments ATLAS and CMS. They have obtained bounds
of up to 1.4~TeV which however depend
on the mass difference between the lighter chargino and the lightest neutralino and to some extent also on
the details of the decays 
\cite{ATLAS:2019lng,ATLAS:2021moa,CMS:2021edw,CMS:2021cox}.
The points below the orange line
feature a chargino mass of above 180 GeV 
and the mass difference to the lightest
neutralino is below 30 GeV, impyling
that these points pass the experimental
bounds as these are higgsino-like states. For the other points a more detailed investigation is required which
we postpone to a future work.

\begin{figure}[H]
\centering
\includegraphics[width=0.8\textwidth]{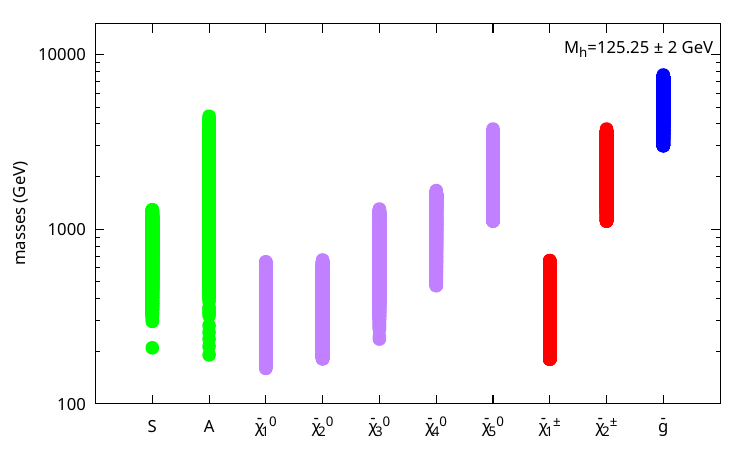}
\caption{\textit{The plot shows  the predicted spectrum for points with Higgs mass within the 2~GeV theoretical uncertainty and lightest chargino mass above 180~GeV. The green points are the CP-even and CP-odd  singlet scalar masses; the purple points are the neutralino masses; the red ones are the chargino masses, followed by the blue points indicating the gluino masses.}}
\label{fig:spectrum}
\end{figure} 

\newpage

The predicted particle spectrum of the model is given in \reffi{fig:spectrum}. The the CP-even singlet scalar is denoted as $S$ while the CP-odd as $A$.
They are not affected by existing searches as they can hardly be produced at the LHC because they are gauge singlets. 
$\tilde{\chi}_i^0, \tilde{\chi}_1^{\pm}$ and $\tilde{g}$ are the neutralinos, charginos and the gluinos, respectively. The points shown
correspond to the ones below the orange
line in \reffi{fig:charneut} to ensure
that they are compatible with existing
searches at the LHC. Adding the other 
points wouldn't change the picture significantly and the most important change would be somewhat smaller values for 
mass of the lightest neutralino.
Note, that the gluino is predicted to be heavier than $2$ TeV in this model
and, thus, this model can explain why so far no sign for supersymmetry has
been found at the LHC. 
This also implies that this model
will be difficult to probe in the coming
LHC runs. The reach of the high luminosity LHC for the lightest
chargino can go up to 200 GeV if the
systematics are very well under control
\cite{Barducci:2015ffa}. Other 
possibilities are the combined 
production of a heavier neutralino 
together with the lightest chargino 
which we will investigate in an upcoming
work. Last but not least we point
out that the lightest neutralino is
an admixture of the singlet fermion and a higgsino and thus it can be a cold
dark matter candidate consistent with
observations, which we will investigate
together with details of the collider 
searches.

\section{Conclusions}\label{conclusions}

We started with a $10D$, $\mathcal{N}=1$, $E_8$ Yang-Mills-Dirac theory with Weyl-Majorana fermions, constructed on the mofified flag manifold $SU(3)/U(1)\times U(1)\times \mathbb{Z}_3$. The CSDR and Wilson flux breaking mechanisms lead to a softly broken $\mathcal{N}=1$, $SU(3)^3\times U(1)^2$ effective  theory in $4D$. Futher breaking of the gauge theory results in a Split NMSSM  scenario that features top, bottom and light Higgs masses  within the experimental limits set by the LHC. The predicted particle spectrum is above the bounds of (so far) non-detection and gives a neutralino LSP with mass lower than $800$GeV.

\subsection*{Acknowledgments}
We would like to thank George Manolakos and Pantelis Manousselis for useful discussions in the theoretical aspects of this work. GP is supported by the Portuguese Funda\c{c}\~{a}o para a Ci\^{e}ncia e Tecnologia (FCT) under Contracts UIDB/00777/2020, and UIDP/00777/2020, these projects are partially funded through POCTI (FEDER), COMPETE, QREN, and the EU. GP has a postdoctoral fellowship in the framework of UIDP/00777/2020 with reference BL154/2022\_IST\_ID.
GP and GZ would like to thank CERN-TH for the hospitality and support. GZ would like to thank the MPP-Munich and DFG Exzellenzcluster 2181:STRUCTURES of Heidelberg University for support.

\bibliographystyle{h-physrev5}
\bibliography{main}

\end{document}